\documentclass[aps,prl,superscriptaddress,reprint,amsmath,amssymb,floatfix,notitlepage,longbibliography]{revtex4-1}
\usepackage[utf8]{inputenc}
\usepackage[T1]{fontenc}
\usepackage{graphicx}
\usepackage{microtype}
\usepackage[usenames,dvipsnames]{xcolor}

\newcommand{\kp}{k_{z}}
\newcommand{\kz}{k_{z}}
\newcommand{\Fext}[1][]{F_{\mathrm{ext}#1}}
\newcommand{\Urad}[1][]{U_{\mathrm{rad}#1}}
\newcommand{\Uelext}[1][]{U^{el}_{\mathrm{ext}#1}}

\begin{document}

\title{Strong coupling between weakly guided semiconductor nanowire modes and an organic dye}

\author{Diego R. Abujetas}
\affiliation{Instituto de Estructura de la Materia (IEM-CSIC), Consejo Superior de Investigaciones Científicas, Serrano 121, 28006 Madrid, Spain}

\author{Johannes Feist}
\affiliation{Departamento de Física Teórica de la Materia Condensada and Condensed Matter Physics Center (IFIMAC), Universidad Autónoma de Madrid, E-28049 Madrid, Spain}

\author{Francisco J. García-Vidal}
\affiliation{Departamento de Física Teórica de la Materia Condensada and Condensed Matter Physics Center (IFIMAC), Universidad Autónoma de Madrid, E-28049 Madrid, Spain}

\author{Jaime Gómez Rivas}
\affiliation{Institute for Photonic Integration, Department of Applied Physics, Eindhoven University of Technology, P.O. Box 513, 5600 MB Eindhoven, The Netherlands}

\author{José A. Sánchez-Gil}
\affiliation{Instituto de Estructura de la Materia (IEM-CSIC), Consejo Superior de Investigaciones Científicas, Serrano 121, 28006 Madrid, Spain}

\date{\today}

\newcommand{\todo}[1]{\textsc{\textcolor{red}{#1}}}

\begin{abstract}
The light-matter coupling between electromagnetic modes guided by a semiconductor nanowire and excitonic states of molecules localized in its surrounding media is studied from both classical and quantum perspectives, with the aim of describing the strong coupling regime. Weakly guided modes (bare photonic modes) are found  through a classical analysis, identifying those lowest-order modes presenting large electromagnetic fields spreading outside the nanowire, while preserving their robust guided behavior. Experimental fits of the dielectric permittivity of an organic dye that exhibits excitonic states are used for realistic scenarios. A quantum model properly confirms through an avoided mode crossing that the strong coupling regime can be achieved for this configuration, leading to Rabi splitting values above 100 meV. In addition, it is shown that the coupling strength depends on the fraction of energy spread outside the nanowire, rather than on the mode field localization. These results open up a new avenue towards strong coupling phenomenology involving propagating modes in non-absorbing media.
\end{abstract}

\maketitle

\section{Introduction}

\begin{figure}
\includegraphics[width=\linewidth]{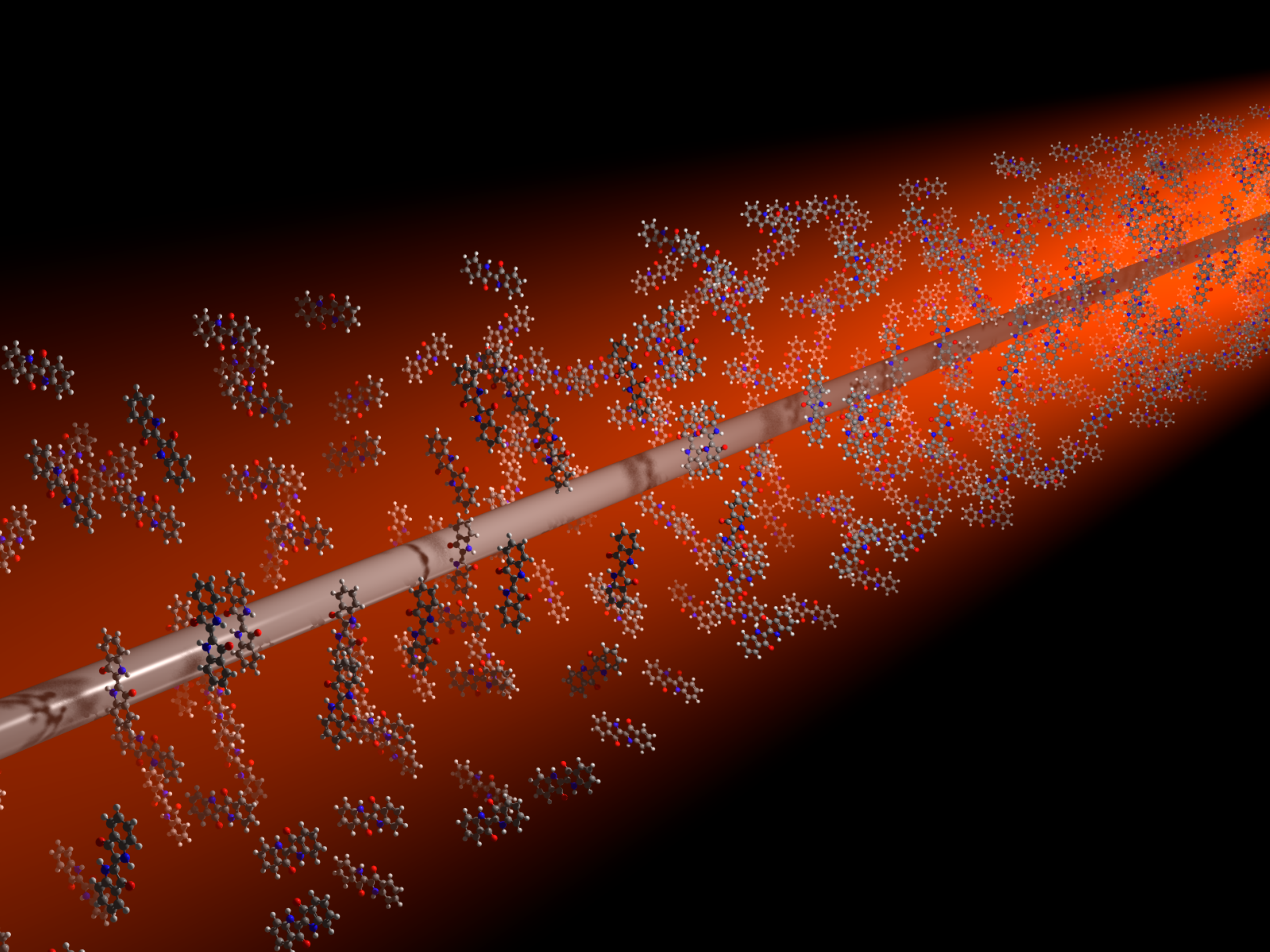}
\caption{Sketch of the setup.}\label{fig:sketch}
\end{figure}

Tailoring light-matter interaction at the nanoscale is the foundation to improve, beyond unpredictable limits, the efficiency of previous devices and to develop new novel applications~\cite{Hutchison2013}. Among others, much effort has been done to engineer the emission properties between electronic energy states of system as quantum dots, wells, and dye molecules, through the coupling to optical systems as cavities, photonic crystals, metallic interfaces and semiconductor wires~\cite{Ghosh2006,Novotny2010,Andreani1999,Zhu1990,Khitrova2006}. Depending on the strength of the coupling between the systems two distinct regimes, weak and strong, can be established. In the weak regime, the spontaneous emission rate is strongly affected by the electromagnetic local densities of states and it can be completely suppressed or enhanced by several orders of magnitude~\cite{Haroche1989,Taminiau2008,Kuhn2006,Yablonovitch1993}, but the natural frequency of the transition remains unaltered. Otherwise, the strong regime is characterized by a coherent exchange of energy between modes inducing new hybrid states with fascinating properties that can be very different from those of the initial systems~\cite{Wallraff2004,Rodriguez2014,Rodriguez2013,Wang2016}.

Metallic nanostructures, through localized surface plasmons (LSP) and surface plasmon polaritons (SPP), can effectively couple to electronic transition states due to their optical near-field enhancement and confinement~\cite{Pockrand1982,Houdre1996,Bellessa2004,Bellessa2009,Gonzalez-Tudela2013,Shi2014,Torma2015}. However, the presence of losses limits their employment in transport applications. In this regard, semiconductor nanowires overcome this issue and allow for a long-rage coupling through propagating guided modes, being in turn a suitable platform to manage the electromagnetic environment at optical frequencies at the nanoscale~\cite{Yan2009}. They possess strong optical resonances and/or guided modes that can be richly tuned by their geometrical and/or material properties~\cite{Yan2009,Abujetas2015,Paniagua-Dominguez2013}. Nevertheless, they have been mainly studied  as optical cavities~\cite{Reithmaier2004,Hennessy2007,Vugt2011,Kuruma2016}, in which quantum dots or wells are placed inside the nanowire during the growing process, constricting light propagation inside. In fact, to the best of our knowledge, coupling nanowire propagating modes to external excitonic media has not been studied yet; this will presumably have a strong impact in exciton transport applications~\cite{High2008,Menke2013,Feist2015,Gonzalez-Ballestero2015}.

In the present work, we study theoretically the appearance of strong coupling regimes in a system consisting of a semiconductor nanowire embedded in an excitonic medium, by means of the interplay between guided modes and excitonic states. Upon exploiting the evanescent tail of various weakly guided modes outside the semiconductor nanowire, analyzed in detail through classical electrodynamics (Sec. II),  coupling to excitonic modes of an organic dye surrounding the nanowire is plausible. A quantum model is developed to properly determine the polaritonic modes revealed through an avoided crossing with expectedly large enough Rabi splittings (Sec. III), showing  that a strong coupling regime can be accomplished. The distribution of the energy is also affected by the coupling (Sec. IV), going from pure photonic to excitonic states, revealing the hybrid nature of the modes. These entangled modes can be relevant for exciton transport purpose, for which the half-life/propagation-length must be optimized (Sec. V).

\section{Weakly guided semiconductor nanowire modes}
We study the dispersion relation of polaritons arising from the coupling between
guided modes in semiconductor nanowires and excitons in a surrounding molecular
medium, through both classical and quantum models. We first discuss the
classical electromagnetism approach, which is based on solving Maxwell's
equations to obtain guided modes in the system. Here, the molecular medium
surrounding the nanowire is modelled through its dielectric function, with
excitons manifesting as resonances leading to broad absorption bands. In a
second step, we discuss a quantum model in which the guided photonic modes of
the semiconductor nanowires are quantized explicitly and coupled to dye
molecules modelled as point dipole emitters~\cite{Gonzalez-Tudela2013}, with a
level structure and molecular density that reproduces the classical dielectric
function in the (linear) low-excitation limit.

For both procedures, the dispersion relation of the nanowire modes must be
solved within classical electromagnetism. Cylindrical waveguides support
propagating waves (leaky and guided) with a dispersion relation determined by:
\begin{multline}
m^{2}\dfrac{\left(k_{z}R\right)^{2}}{\left(\omega R/c\right)^{2}} = 
  \left[ \dfrac{\mu_{c}}{u}\dfrac{J'_{m}(u)}{J_{m}(u)} - \dfrac{\mu_{b}}{v}\dfrac{H'_{m}(v)} {H_{m}(v)} \right] \times \\
  \left[\dfrac{\varepsilon_{c}}{u}\dfrac{J'_{m}(u)}{J_{m}(u)} - \dfrac{\varepsilon_{b}}{v}\dfrac{H'_{m}(v)}{H_{m}(v)} \right],
\label{eq:RD}
\end{multline}
where $R$ is the radius of the cylinder, $\omega$ is the angular frequency,
$\kz$ is the wavevector of the mode along the cylinder axis, and $m$ an integer
related with the azimuthal distributions of the fields. Furthermore, $J_{m}$ and
$H_{m}$ are the Bessel and Hankel functions of the first kind, and $'$ denotes
the derivative with respect to the argument. The parameters $u=k_{c}R$ and
$v=k_{b}R$ are proportional to the transverse component of the wave vector
inside and outside the cylinder, respectively, given by
\begin{subequations}
\begin{align}
k_{c}^{2} &= \varepsilon_{c}\mu_{c}\dfrac{\omega^{2}}{c^{2}} - k_{z}^{2}, \\
k_{b}^{2} &= \varepsilon_{b}\mu_{b}\dfrac{\omega^{2}}{c^{2}} - k_{z}^{2},
\end{align}
\end{subequations}
where $c$ is the speed of light in vacuum, while $\varepsilon_{c}$, $\mu_{c}$
and $\varepsilon_{b}$, $\mu_{b}$ are the electric permittivity and magnetic
permeability of the cylinder and the background medium, respectively. In the
following, we use $\mu_c=\mu_b=1$.

\begin{figure}[tb]
  \includegraphics[width=\linewidth]{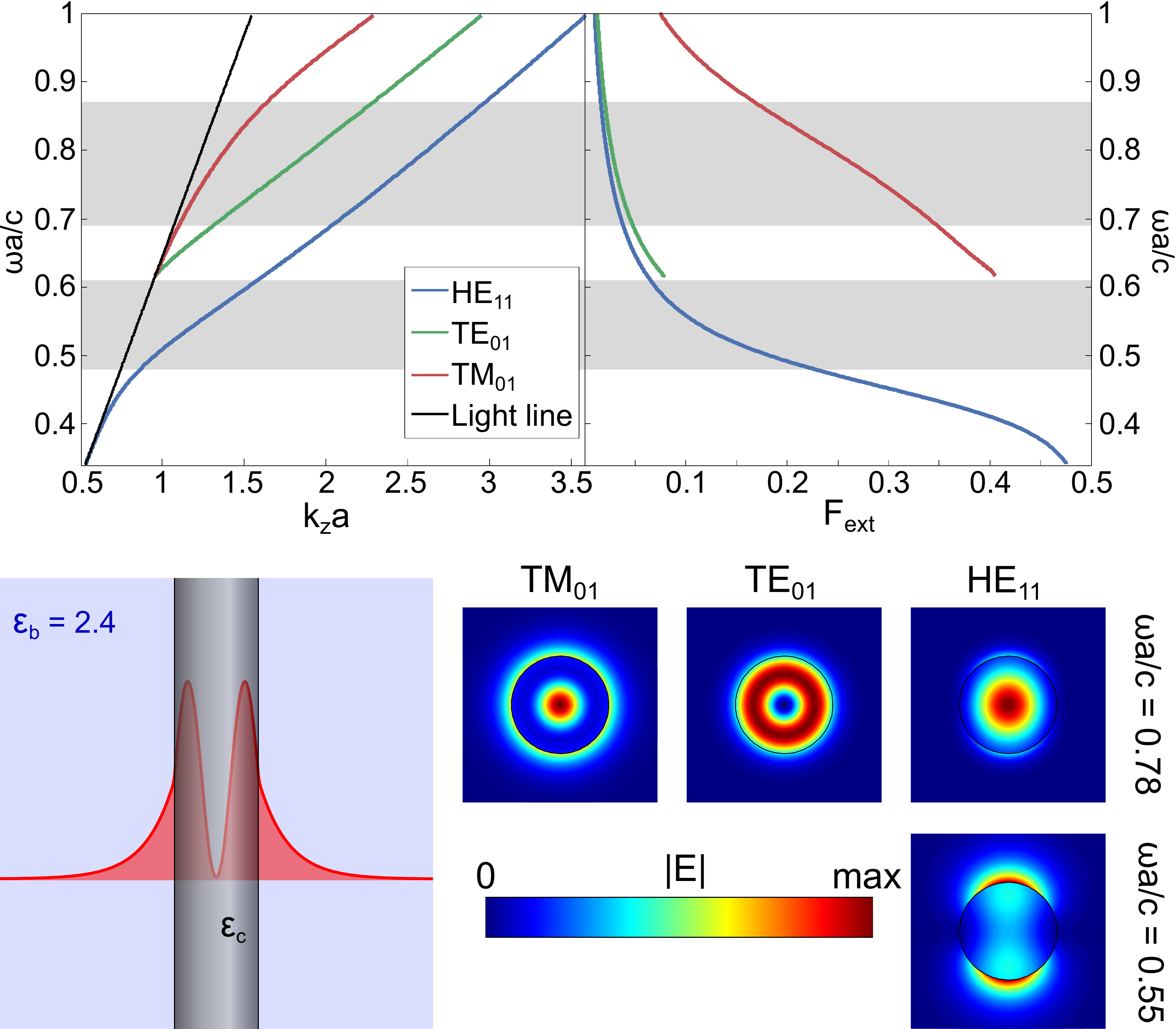}
  \caption{Bare system. Top panel: Dispersion relation $\omega R/c$ vs
  $k_{z}R$ for the first three guided modes for $\varepsilon_{b} = 2.4$ and
  $\varepsilon_{c} = 4.2^{2}$. Bottom panel: Schematic of the waveguide system and
  norm of the electric field profile of each mode at $\omega R/c = 0.55$ and
  $\omega R/c = 0.78$.}\label{fig:bare}
\end{figure}

Recall that for each subindex $m$ on Eq.~\ref{eq:RD} there are associated several solutions that can be denoted by the subscript $l$. Hence, a pair index $ml$ can be associated to each guided mode. For guided modes with $m=0$, the field is symmetric about the cylinder axis, exhibiting a pure transverse character, either electric (TE$_{0l}$, $E_z = E_r = H_\phi = 0$) or magnetic (TM$_{0l}$, $H_z = H_r = E_\phi = 0$). Hybrid modes arise for $m \not = 0$ (HE$_{ml}$), where in general all field components are non-zero and their phases accumulate a factor of $2\pi m$ in a closed loop around the cylinder axis.

We first consider the ``bare'' modes of the nanowire embedded in a host material
without organic molecules. We consider high-refractive-index, lossless
semiconductor nanowires; without loss of generality, a refractive index of $n =
\sqrt{\varepsilon_c} = 4.2$ (close to those of GaAs, GaP, AlSb in the visible)
is used for the nanowire, while the background dielectric constant is set to
$\varepsilon_b=\varepsilon_h=2.4$ (typical for polymers such as PMMA or PVAc).
The upper left panel of Fig.~\ref{fig:bare} shows the dispersion relations ($k_z
R$ vs.\ $\omega R/c$) of the first three guided modes (HE$_{11}$, TE$_{01}$,
TM$_{01}$). In order to optimize the coupling of these modes to molecules that
will be placed in the medium surrounding the nanowire, the mode should carry as
much energy as possible outside the wire. This implies that the maximum coupling
can be achieved with weakly guided modes close to the light line, since their
field profiles possess large evanescent tails outside the wire. To quantify
this, the upper right panel of Fig.~\ref{fig:bare} shows $\Fext$,
defined as the fraction of mode energy stored in the electric field outside the
nanowire. As the excitonic transitions of dye molecules correspond, to a very
good approximation, to electric dipole transitions, only the energy density from
the electric field is taken into account. This implies that $\Fext \leq
0.5$, since for guided propagating modes the energy is equally divided between
electric and magnetic fields. As can be seen, the cutoff-free HE$_{11}$ mode at
lower frequencies and the (predominant) transverse magnetic modes (for
dielectric waveguides) close to their cut-off frequencies are both candidates to
exhibit strong coupling phenomenology, $\Fext$ coming close to its
maximum value of $0.5$, and both modes becoming more bounded as the normalized
frequency $\omega R/c$ increases. For transverse electric modes, the continuity
of all field components across the boundaries causes a flatter dispersion
relation and a larger confinement of the field inside the nanowire. As we will
see later, strong coupling can still be achieved for TE modes, albeit with
smaller Rabi splittings. The bottom of Fig.~\ref{fig:bare} shows the electric
field intensity profiles for each mode at normalized frequencies $\omega R/c =
0.55$ (where only the HE$_{11}$ mode exists) and $\omega R/c = 0.78$, as well a
pictorial representation of the system. The field profiles confirm the
information of $\Fext$, showing large electromagnetic fields outside
the nanowire for the HE$_{11}$ and TM$_{01}$ modes. We note that in the upper
panels of Fig.~\ref{fig:bare}, the shaded areas mark the spectral regions where
the excitonic states are located for nanowire diameters $D = 2R = 90$, $120$,
and $130$~nm, as will be studied below.

\begin{figure}[tb]
  \includegraphics[width=\linewidth]{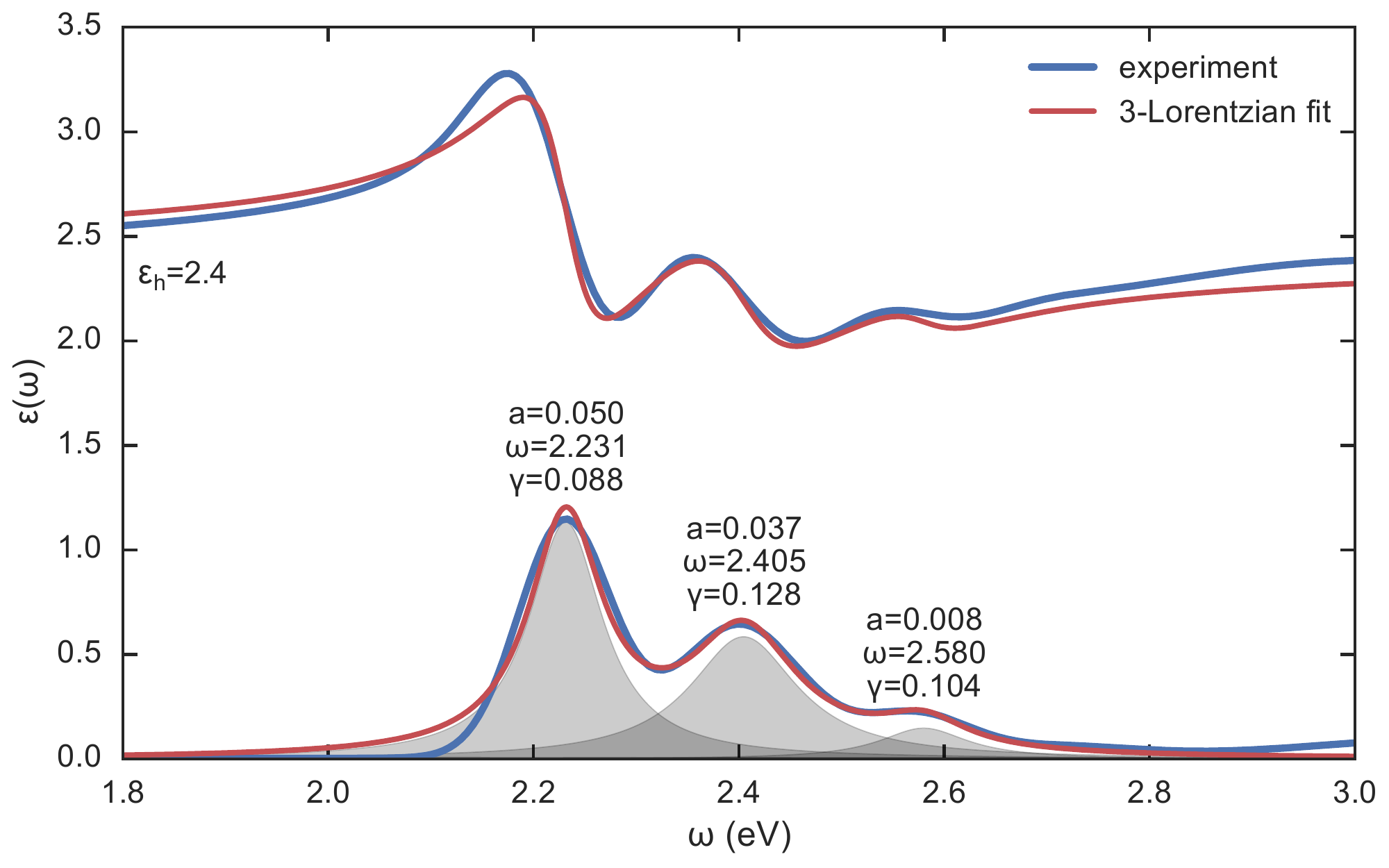}
  \caption{Fit of Lorentzians to the New Pink dielectric function (real and imaginary parts), determined experimentally.}\label{fig:eps_np}
\end{figure}
  
After identifying the suitable bare photonic modes of the system, we now include
the effect of an organic dye within the host medium, first staying within a
classical description. Without loss of generality, we choose ``New Pink'' as the
dye molecule. This molecule has been used in various experiments achieving
strong coupling as it shows little biexciton annihilation even at high
densities, and is well-characterized~\cite{Ramezani2017Plasmon,
Ramezani2018Dispersion}. Its measured electric permittivity (both real and
imaginary part) is shown in Fig.~\ref{fig:eps_np} (blue lines), together with a
fit to a model dielectric function containing three Lorentzian resonances to
represent dye excitations (red lines):
\begin{equation}
\varepsilon(\omega) = \varepsilon_{h} + \sum_{k=1}^3 \frac{a_k}{\omega_k - i \gamma_k/2 - \omega},
\label{eq:dielectric}
\end{equation}
where $\varepsilon_{h}$ is the background permittivity of the host medium and
$\omega_k$, $\gamma_k$ and $a_k$ are the frequency, decay rate, and amplitude of
each resonance, respectively. The fit parameter values are given in
Fig.~\ref{fig:eps_np} next to each peak. While the fit with Lorentzian resonances
is reasonably accurate, we note here that a nearly perfect fit to the dielectric
function can be achieved by using Voigt profiles instead of Lorentzian ones.
These correspond to the convolution of Lorentzians with Gaussians, and can
represent both homogeneous and inhomogeneous broadening, while only homogeneous
broadening (i.e., losses and dephasing) is accounted for through Lorentzian
profiles. 

As the physical results do not change significantly (we have compared both
approaches), for simplicity we use the Lorentzian fit to calculate the
dispersion relations. However, since Lorentzians have much longer tails than
seen in the experimental absorption spectrum, this approximation significantly
overestimates the losses at frequencies below about $2.1$~eV. As we will see
later, using the experimental dielectric function (or equivalently, the fit to
Voigt profiles) leads to significantly longer lifetimes and propagation lengths
when strong coupling ``pushes'' the polaritonic states away from the molecular
resonances.

\begin{figure*}[tb]
\includegraphics[width=\linewidth]{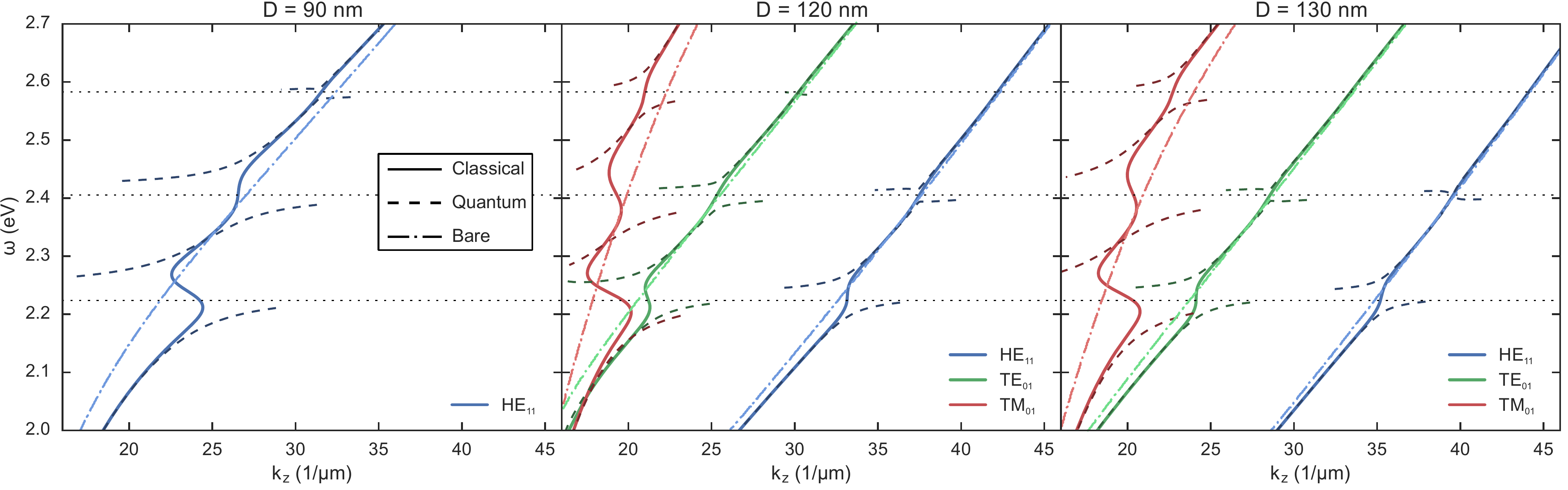}
\caption{Comparison between classical electromagnetics calculation of mode
  dispersions (solid curves) from Eq.~\ref{eq:RD} with $\varepsilon_{c}=4.2^{2}$
  and $\varepsilon_{c}$ from the fit to Eq.~\ref{eq:dielectric} shown in
  Fig.~\ref{fig:eps_np}, and quantum model (dashed curves), Eq.~\ref{eq:Ha}, for
  three nanowire diameters $D=90$, $120$, $130$ nm. For clarity, only quantum
  modes with significant photon fraction are shown (the parts not shown are
  basically straight lines at the bare emitter frequencies, marked by dotted
  horizontal lines). For comparison, the bare dispersion relations are shown as
  dot-dashed curves (Eq.~\ref{eq:RD} with $\varepsilon_{c}=4.2^{2}$ and
  $\varepsilon_{c} = 2.4$)}\label{fig:model_dispersion}
\end{figure*}

The dispersion relation of the nanowires surrounded by molecules, as calculated
within a classical approach by solving Fig.~\ref{eq:RD}, is shown in
Fig.~\ref{fig:model_dispersion} for the three first modes (solid curves) and for
three different wire diameters $D = 90, 120, 130$~nm, in order to analyze the
coupling to different modes. Here, the dot-dashed curves represent the
dispersion relations for the bare system and the (horizontal) dotted lines
represent the resonance frequencies of the excitons. In contrast to the
bare-wire case, the lossy nature of the dye resonances implies that the wave
vector $k_z$ acquires an imaginary part representing the propagation losses of
the polariton modes. The dispersion relations show significant energy shifts
close to the resonances of the molecule excitons, and also feature a back-bending
that can indicate a mode hybridization, i.e., strong coupling or polariton
formation, in the classical calculations. As expected, the
observed splitting is more pronounced for bare photonic modes that are only
weakly confined within the nanowire, as well as for dye resonances with larger
associated transition dipole moments (i.e., larger absorption amplitude $a_n$).
However, it should be noted that even the more strongly confined TE$_{01}$ mode
displays back-bending at the first excitonic resonance.

\section{Quantum model: Rabi splittings}

The classical analysis shows bending bands in the dispersion
relation, which are an indicative signature of a strongly coupled system in which
avoided crossings arise at resonances. Nonetheless, the real coupling is difficult to
quantify without a representative quantity such as a Rabi splitting, which has a
direct meaning in a quantum description, but does not show up explicitly in the
classical calculation.

To construct a quantum model, we proceed in a similar manner as
in~\cite{Gonzalez-Tudela2013}. We start by quantizing the guided bare-nanowire
modes by placing the system within a box of length $L$ along the wire axis and
imposing periodic boundary conditions in this direction. This restricts the
allowed values of the parallel momentum to $\kz=\frac{2\pi n}{L}$, with
$n\in\mathbb{Z}$. Since the bare nanowire modes are lossless and confined in the
transverse direction, this also allows for their straightforward quantization by
imposing that the integrated energy density is equal to the photon energy (see,
e.g., the appendix of~\cite{Gonzalez-Ballestero2015}). Defining the quantized
field profile $\vec{\mathcal{E}}(\vec r) = C \vec E(r) e^{i \kz z + i m \phi}$
in cylindrical coordinates $r,\phi,z$, where $\vec E(r)$ is the electric field
profile of the mode with arbitrary normalization, gives
\begin{equation}
  C = \sqrt{\frac{\hbar\omega}{2\pi L \Urad}}, 
\label{eq:norm}
\end{equation}
where $\Urad$ is an integral over the electromagnetic energy density of the
mode, given by
\begin{equation}\label{eq:Urad}
\Urad = 2 \int \epsilon(r) |\vec E(r)|^2 \mathrm{d}r.
\end{equation}
Here, the factor $2$ accounts for the fact that equal energy is stored in the
magnetic field and in the electric field. Note that we are only treating guided
waveguide modes, for which $\Urad$ is well-defined as $k_b$ is purely imaginary
and the mode profile decays exponentially far away from the wire. In addition,
there is a continuum of freely propagating modes inside the light cone, which we
neglect as they do not play a large role in the situations we study here
(although they can have important effects in specific
cases~\cite{Yuen-Zhou2017}). The Hamiltonian of the system within the
rotating-wave approximation is then given by
\begin{multline}\label{eq:Hquant}
  H = \sum_{n,m} \omega_{c,nm} \hat{a}_{nm}^\dagger \hat{a}_{nm} + \sum_j H_{\mathrm{mol},j}\, + \\
      \sum_{n,m,j} \vec{\mathcal{E}}_{nm}(\vec r_j) \cdot \vec d_j \, (\hat\mu_j^- \hat{a}_{nm}^\dagger + \hat\mu_j^+ \hat{a}_{nm}),
\end{multline}
where $\hat{a}_{nm}$ is the bosonic annihilation operator corresponding to the
$m$th mode with parallel momentum $\kz=\frac{2\pi n}{L}$, while
$H_{\mathrm{mol},j}$ and $\hat\mu_j\!=\!\hat\mu_j^+\!+\!\hat\mu_j^-$ are the bare
Hamiltonian and dipole operator of molecule $j$, respectively, and $\vec d_j$ is
a unit vector describing the orientation of the molecule. We have here neglected
an extra term (proportional to $A^2$ or $\hat\mu^2$, depending on gauge) in the
light-matter interaction, which only becomes important in the limit of
ultrastrong coupling, i.e., when coupling strengths become comparable to the
bare transition frequencies~\cite{Ciuti2006,
DeLiberato2017,DeBernardis2018Cavity,Schafer2018}. The dye molecules are
represented as few-level emitters with parameters chosen to reproduce the
macroscopic dielectric function. In particular, we treat the molecules as
four-level systems, with one ground and three excited states,
\begin{align}
H_{\mathrm{mol},j} &= \begin{pmatrix}
  0 & 0 & 0 & 0\\
  0 & \omega_1-i\frac{\gamma_1}{2} & 0 & 0\\
  0 & 0 & \omega_2-i\frac{\gamma_2}{2} & 0\\
  0 & 0 & 0 & \omega_3-i\frac{\gamma_3}{2}
\end{pmatrix},\\
\hat\mu_j &= \begin{pmatrix}
  0   & a_1 & a_2 & a_3\\
  a_1 & 0 & 0 & 0\\
  a_2 & 0 & 0 & 0\\
  a_3 & 0 & 0 & 0
\end{pmatrix},
\end{align}
where the parameters $\omega_k$, $\gamma_k$, and $a_k$ are taken from the fit in
Eq.~\ref{eq:dielectric}. Note that we also neglect direct dipole-dipole
interactions between the molecules, as their (averaged) effect is already
included in the transition frequencies $\omega_k$ extracted from the dielectric
function. We note for completeness that an alternative (but much more costly)
approach would be to extract the molecular parameters from a fit to the
bare-molecule polarizability (obtained from the dielectric function using the
Clausius-Mossotti relation), and then explicitly include dipole-dipole
interactions between the molecules.

We treat the experimentally relevant limit that the host material contains many
randomly oriented organic dye molecules, distributed evenly in the region around
the nanowire with number density $\rho_{\mathrm{mol}} = 1/V_{\mathrm{mol}}$,
where $V_{\mathrm{mol}}$ is the average volume occupied by each molecule.
Considering the random distribution of the molecules along the wire,
translational symmetry is approximately conserved~\cite{Gonzalez-Tudela2013},
and, consequently, superpositions of molecular states can be formed with a
well-defined wavevector $k_z$. 
The Hamiltonian thus becomes (approximately) diagonal as a function of the
parallel wavevector index $n$, significantly simplifying its diagonalization. 




In addition, for the case that more than a single guided mode exists at a given
$\kp$ (as is the case for $D=120$\,nm and $D=130\,$nm), we can also approximate
that since the nanowire modes are orthogonal, they couple to independent Dicke
states (superpositions of molecular excitations). This implies that the coupling
between different wire modes and molecular excitations is independent. The
collective coupling strength between the $i$th mode with parallel momentum $k_n$
and the molecular Dicke state corresponding to the $k$th excitation is then
given by
\begin{equation}
g_{nmk}^2 = \sum_{j} |\vec{\mathcal{E}}_{nm}(\vec r_{j}) \cdot \vec d_{j} a_k|^2
\end{equation}
Considering the cylindrical symmetry of the system, and that the molecules are randomly oriented and fill all of space outside the wire evenly, the coupling strength can be approximated by 
\begin{equation}
g_{nmk}^2 \approx \frac{4\pi L}{3} \frac{a_k^2}{V_\mathrm{mol}} \int_R^\infty r |\vec{\mathcal{E}}_{nm}(r)|^2 \mathrm{d}r.
\label{eq:gn1}
\end{equation}
Using Eq.~\ref{eq:norm} and Eq.~\ref{eq:Urad}, the coupling strength can be
written as
\begin{equation}
g_{nmk}^2 = \frac{2\hbar\omega_{c,nm}}{3\pi \epsilon_{h}} \frac{a_k^2}{V_\mathrm{mol}} \frac{\Uelext[,nm]}{\Urad[,nm]},
\end{equation}
where $U^{el}_{ext}$ is the part of the mode energy stored in the electric field
outside the wire
\begin{equation}
U^{el}_{ext}= \int_{r>R} \epsilon_{h} |\vec{\mathcal{E}}(\vec r)|^2  \mathrm{d}\vec r.
\end{equation}
This can be expressed through $\Fext = \Uelext/\Urad$ and the molecular density
$\rho_\mathrm{mol}$ as
\begin{equation}
  g_{nmk} = a_k \sqrt{\frac{2\hbar\rho_\mathrm{mol}}{3\pi \epsilon_{h}} \omega_{c,nm} \Fext[,nm]}.
  \label{eq:gfinal}
\end{equation}
It is interesting to note that the coupling strength does not depend on whether
the guided nanowire mode has a small mode volume (strong localization of the
field). The only information about the wire mode entering the final expression
is its frequency and the fraction of the mode energy that is outside the wire.
Note that this assumes that space is completely filled with molecules, so that a
less confined mode effectively interacts with more molecules to give the same
(or even larger) collective coupling as a confined mode. The localization of the
mode is compensated by the field strength and in that sense, well-confined (out
the wire) modes are only advantageous in terms of needing less space, but not in
terms of reaching strong coupling.

We can now take the limit $L\to\infty$, such that $\kp=2\pi n/L$ becomes a
continuous variable, and proceed to construct an effective $4 \times 4$ model
Hamiltonian for each $\kp$ and each nanowire mode independently:
\begin{equation}
H_i(\kp) = \begin{pmatrix}
\omega_{c,m}(\kp) & g_{m1}(\kp) & g_{m2}(\kp) & g_{m3}(\kp)\\
g_{m1}(\kp) & \omega_1-i\frac{\gamma_1}{2} & 0 & 0\\
g_{m2}(\kp) & 0 & \omega_2-i\frac{\gamma_2}{2} & 0\\
g_{m3}(\kp) & 0 & 0 & \omega_3-i\frac{\gamma_3}{2}
\end{pmatrix},
\label{eq:Ha}
\end{equation}
where $m$ labels the nanowire mode, and $g_{mk}(\kp)$ is given by
Eq.~\ref{eq:gfinal}. 

In Fig.~\ref{fig:model_dispersion}, the dispersion relations calculated by
solving Eq.~\ref{eq:Ha} are shown (dashed colored curves) with proper avoided
crossings appearing near excitonic frequencies manifesting the strong coupling
leading to polaritonic modes. Out of resonance, as expected, the dispersion
relations are practically the same as those of the classical bare photonic
modes. It should be pointed out that the classical electromagnetic calculations
work at real energies $\omega$, but allow complex momenta $\kp$ (describing
propagation loss). On the other hand, the quantum model works at real $\kp$, but
allows complex energies (describing temporal loss). These two pictures represent
the same physics, but are not completely equivalent. 
These differences are the reason for the different behavior close to the regions
of largest absorption where the classical modes bend backwards (stationary
electromagnetic field solutions), while the quantum modes are actually split
(time dynamic driving process).  We also note that the good agreement between
quantum and classical calculations in Fig.~\ref{fig:model_dispersion}, without
any fit parameters, cannot be reproduced by the often-used strategy of
constructing an approximate model with $\kp$-independent couplings $g_{mk}$
(corresponding to approximating $\omega_{c} \Fext$ as constant).
 
Furthermore, we emphasize that the quantum model here clearly shows that strong
coupling is reached for these conditions. For example, the cleanest system is
given by the nanowire with diameter $D=90\,$nm, for which only the HE$_{11}$
mode is guided. This system supports strong coupling with significant Rabi
splittings $\Omega_R>100\,$meV for both the first and second molecular
excitations. Furthermore, nanowires with larger diameter support multiple
polaritons (also with significant Rabi splitting) at the same frequency, which
might be interesting for possible applications.

\section{Energy distribution: photonic and excitonic modes}

An important characteristic of strong coupling is coherent energy exchange
between different physical systems, inducing hybrid states that no longer can be
seen/described as individual systems. The distribution of the energy into
photonic and excitonic parts is important to characterize the new states.

From a classical perspective, the electromagnetic energy for lossy media the can
be calculated as a perturbation from the lossless case. However, the strong
dispersion and the high losses of the permittivity invalidate the usual
expressions for the electromagnetic energy stored in the system. In
addition, there is no a clear distinction between the energy stored in the
medium and in the electromagnetic field. For this purpose, we follow the
approach taken by Loudon~\cite{Loudon1970} for the energy of a medium with a
single resonant frequency, and its extension to multiple
resonances~\cite{Oughstun1988,Vazquez-Lozano2018}, considering each excitonic
state as an independent resonance.

The energy of an absorbing dielectric medium described by resonances is
\begin{align}
W &= W_{o} + W_{f}, \nonumber \\
W_{o} &= \dfrac{1}{4}\int_{R}^{\infty} r \varepsilon_{0} |\vec E(\vec r)|^{2}  \mathrm{d}r \sum_{n}{ \dfrac{2a_{n}\omega \left(\omega_{n}^{2} + \omega^{2} \right) }{\left(\omega_{n}^{2} - \omega^{2}\right)^{2} + \omega^{2}\gamma_{n}^{2}}} ,  \nonumber \\ 
W_{f} &= \dfrac{1}{4}\int_{0}^{\infty} r\left[ \varepsilon_{0}\varepsilon_{r} |\vec E(\vec r)|^{2} + \mu_{0}\mu_{r}|\vec H(\vec r)|^{2} \right] \mathrm{d}r, 
\label{eq:Energy}
\intertext{where}
\varepsilon_{r} &= \left\lbrace\begin{array}{c} \varepsilon_{c}, \quad r<R \\
                                               \varepsilon_{h}, \quad r>R \end{array} \right.
\qquad
\mu_{r} = \left\lbrace\begin{array}{c} \mu_{c}, \quad r<R \\
                                       \mu_{b}, \quad r>R \end{array} \right.
.
\end{align}
The first term of Eq.~\ref{eq:Energy}, $W_{o}$, is the energy stored in the
excited oscillators (excitons) and it goes to zero out of the resonances. The
second term $W_{f}$ is the energy carried by the electromagnetic field.

Within the the quantum model, the excitonic energy is calculated as the (real
part of the) expectation value of the molecular exciton Hamiltonian
$H_\mathrm{mol}$ in each state, relative to the total energy of the eigenstate.

\begin{figure}[tb]
\includegraphics[width=\linewidth]{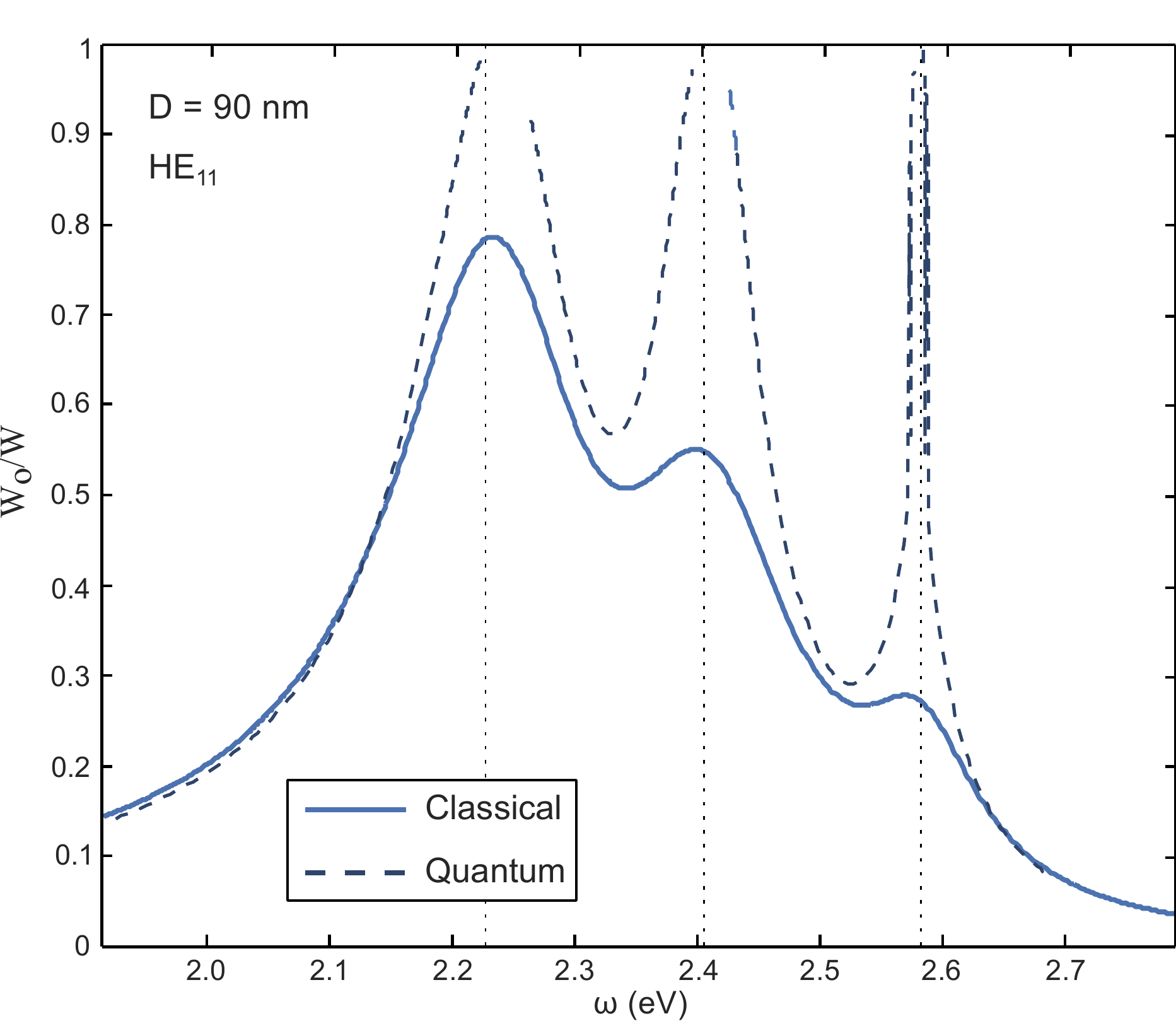}
\caption{Fraction of the energy stored in the excitons calculated by the
classical (solid curve) and quantum approach (dashed curve) for the HE$_{11}$
mode at a nanowire diameter of $D=90$ nm. The material properties are the same
as in Fig.~\ref{fig:model_dispersion}. The resonant frequencies of the excitons
are marked by dotted vertical lines.}\label{fig:Energy}
\end{figure}

As we have previously seen that each nanowire mode is independent from the point
of view of the coupling with the excitonic media, we now focus only on the case
$D=90$~nm (results for other modes are analogous). The ratio between the energy
stored in the oscillators and the total energy of the system, $W = W_{o} +
W_{f}$, characterizes the nature of the mode (photonic/excitonic).
Fig.~\ref{fig:Energy} shows the percentage of the energy stored in the excitons
obtained by the classical (solid curve) and quantum approaches (dashed curve)
for the HE$_{11}$ mode at $D = 90$~nm. The agreement between both approaches is
extremely good at the frequencies where both dispersion relations coincide (see
Fig.~\ref{fig:model_dispersion}). Far from resonance, the energy stored in the
oscillators goes to zero, thus there is no interaction with the excitons and the
mode is practically photonic. In the spectral region of the resonance bands, the
fraction of the energy in the excitons increases, as the mode becomes
polaritonic. Close to resonance, as expected, both approaches differ: whereas
the energy fraction within the classical model yields smooth maximal values
(below 100\%) at the excitonic frequencies (corresponding to the regions in
which the mode dispersion relation bends backwards in
Fig.~\ref{fig:model_dispersion}), the quantum approach asymptotically tends to
one, namely, to 100\% of energy stored as excitons (flat dispersion relation in
Fig.~\ref{fig:model_dispersion}).

It is interesting to note that a high enough fraction of excitonic energy is
required for this system to be a suitable platform for excitonics applications;
however, at the same time, it is desirable to minimize losses, which is achieved
for high photon components. A good compromise can be achieved at intermediate
energy fractions below the onset of losses in the dielectric function ($\omega
\approx 2.1$ eV, at which $W_{o}/W \approx 0.3$), as we will show below.

\section{Half-life, propagation length and energy velocity}

While up to now, we have focused on the real part of the quantities (wave
vectors and/or energies), the losses of the system are also a very important
part necessary to obtain a complete characterization. In particular, they
determine dynamic properties such as the mode propagation length that are
crucial for practical applications. Thus far, we have studied the properties of
the system using a Lorentzian fit for the dielectric function, rather than the
experimentally measured one, since this allowed more direct comparison with a
simple quantum model, and gives almost identical results for the dispersion
relation and to determine the existence for strong coupling. However, the
Lorentzian fit severely overestimates losses away from the resonances, and much
more reliable results are obtained when using the actual experimental values in
order to calculate properties such as the half-life $\tau$ and the propagation
length $L_p$ (linked through the group velocity $v_g$).

\begin{figure}[tb]
\includegraphics[width=\linewidth]{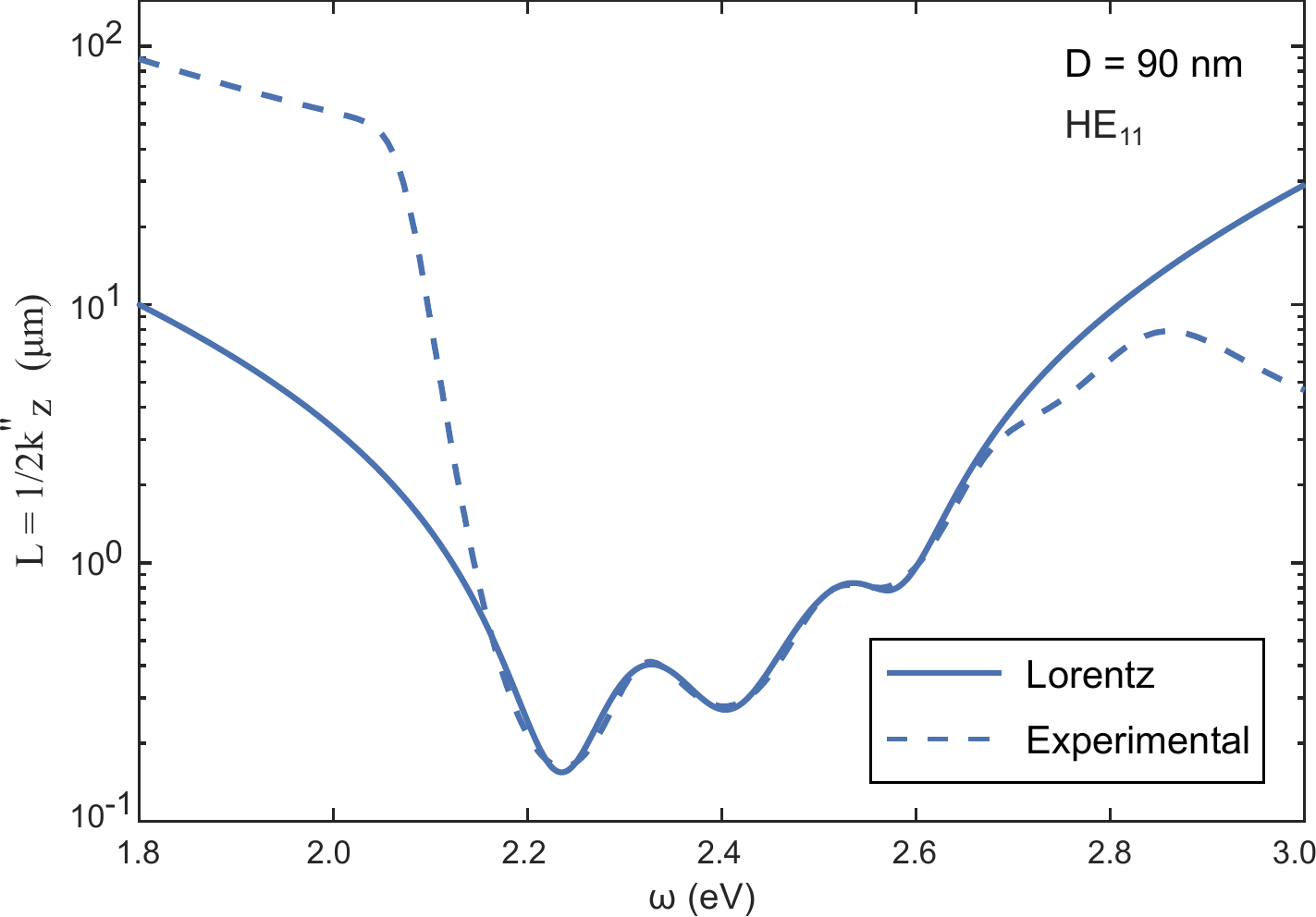}
\caption{Classical calculations of the propagation length $L_{p}$ (in
logarithmic scale) as a function of frequency for the HE$_{11}$ mode
($D=90$~nm). Solid curve: the dielectric function is fitted to a Lorentzian
model, Fig.~\ref{fig:eps_np}. Dashed curves: the experimental value of the
dielectric function is used.}\label{fig:PL}
\end{figure}

The propagation length, $L_{p}$, is the distance after which the energy of the
wave is reduced to $1/e$ of its initial value. For the classical approach, the
propagation length of a guided mode is given by $L_{p} = 1/2\kp''$, inversely
proportional to the imaginary part of the propagation constant, $\kp = \kp' + i
\kp''$, that accounts for the attenuation of the mode. In Fig.~\ref{fig:PL} the
propagation length of the HE$_{11}$ mode for $D=90$~nm (as in
Fig.~\ref{fig:Energy}) is plotted in logarithmic scale as a function of the
frequency, using both the experimental value of the dielectric function (solid
curves) and the Lorentzian fit from Eq.~\ref{eq:dielectric} shown in
Fig.~\ref{fig:eps_np} (dashed curves).

At frequencies below first resonance, $L_{p}$, using the experimental dielectric
function gives values that are typically one order of magnitude larger than with
the Lorentzian fit, and gives propagation lengths that are $1-2$ orders of
magnitude larger than the wavelength of the mode.

We also note that the difference between dielectric functions is much smaller
for the TE$_{01}$ and TM$_{01}$ modes at very low frequencies (not shown here).
This is due to the fact that, as the modes become leaky, the radiation processes
begin to dominate losses and the imaginary part of the permittivity is no longer
determinant. On the other hand, as expected, both methods give similar results
at frequencies close to resonance ($2.2$-$2.6$~eV), where the Lorentzian fits
accurately reproduce the dielectric function. Finally, the Lorentzian fit leads
to higher (unphysical) values of $L_{p}$ at high frequencies since another
resonance appears that was not included in the fit therein.

For current and future applications that rely on exciton
transport~\cite{High2008,Ballarini2013,Menke2013,Feist2015}, large propagation
lengths are desired while simultaneously keeping a considerable amount of energy
in the excitons (excited oscillators). To achieve better coupling, the mode
energy must be outside the wire, although the localization of the energy in the
excitonic medium would imply higher losses. These two issues must be balanced in
order to optimize the propagation length. An important effect to take into
account here is that the propagation length is determined by the dielectric
losses at the frequency of the polariton mode. This implies that strong coupling
can be used to ``push away'' the exciton peak from the exciton losses through
polariton formation, allowing to create states with high exciton character that
do not suffer from the large excitonic losses. This is a particular strength of
the semiconductor nanowire systems studied here, where the bare photonic modes
are essentially lossless. For example, focusing on the HE$_{11}$ mode and
demanding that 30\% of the energy be in the excitons, the maximum propagation
length is reached at a diameter $D = 70$~nm, with $L_{p} = 48~\mu$m at $\omega =
2.02$~eV (at the lower part of the first resonant frequency). The propagation
wavelength of the mode at that frequency is $367~nm$, $130$ times smaller than
the propagation length. As a general consideration, the propagation length is
optimum (related to the energy stored in the excitonic media) close to
frequencies at which losses start to substantially increase due to inhomogeneous
broadening.

The velocity at which energy is transported by a propagating mode is normally
given by the group velocity, $v_{g}$, defined as
\begin{align}
v_{g} = \frac{\mathrm{d}\omega}{\mathrm{d}k}.
\label{eq.vga}
\end{align}
However, the velocity at which the energy is transported must always be lower
than the speed of light, which is not fulfilled by this definition; for example,
in the region where the photonic bands bend backward, the group velocity goes to
infinity. Therefore, as for the energy, an alternative definition is needed in
lossy media. Following the works of Loudon and Brillouin~\cite{Loudon1970,
Brillouin1960}, we define the energy velocity, $v_{e}$, as the ratio between the
flow of energy, given by the Poynting vector, and the total energy stored in the
system
\begin{align}
v_{e} = \frac{S_{z}}{W} = \frac{\int_{0}^{\infty}rS_{z}(\vec r)dr}{W}.
\label{eq.vgb}
\end{align}
For the bare photonic states calculated before (lossless case), the results
given by Eq.~\ref{eq.vga} and Eq.~\ref{eq.vgb} are identical (not shown here).

\begin{figure}[tb]
\includegraphics[width=\linewidth]{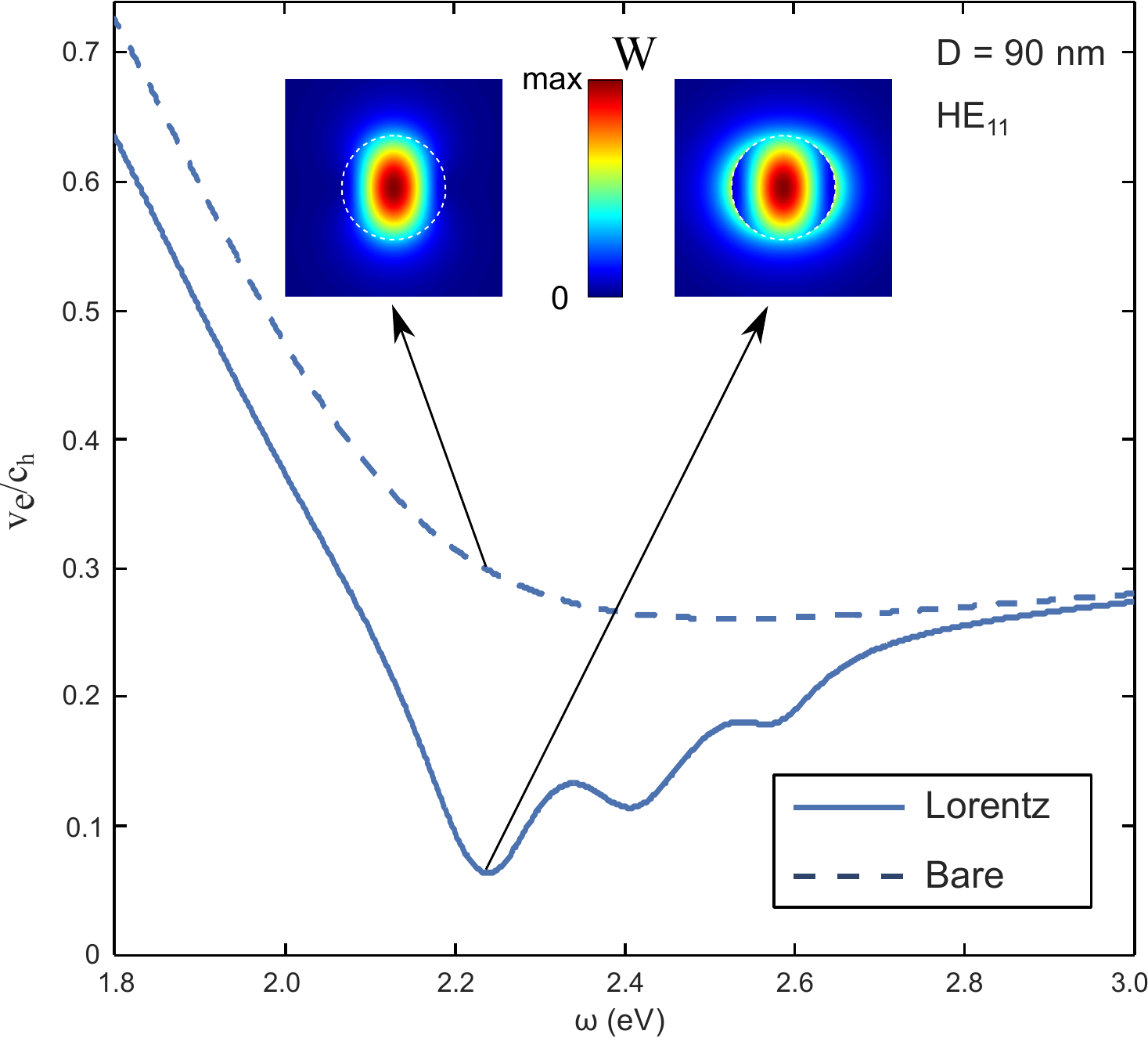}
\caption{Classical calculations of the energy velocity $v_{e}$ divided by the speed of light in the host medium, $c_{h}$, as a function of  frequency for the HE$_{11}$ mode ($D=90$~nm): Solid curve: dielectric function fitted to a Lorentzian model; dashed curve: bare system ($\varepsilon_{b} = 2.4$). The insets show the energy distribution at the first resonant frequency $\omega = 2.23$~eV for both cases.}\label{fig:VG}
\end{figure}

In Fig.~\ref{fig:VG} the energy velocity (divided by the speed of light in the
host medium, $c_{h}$) of the HE$_{11}$ mode is shown for $D = 90$~nm. In
addition, the result for the bare system is also included (dashed line). The
energy velocity presents dips at resonant frequencies. Strong coupling manifests
as a reduction in the energy velocity, since the coherent exchange of energy
slows down the light flow. At resonant frequencies, the electric permittivity of
the excitonic media matches the value of the background $\varepsilon_{h}$.
Despite the non-zero imaginary part, the field profiles for both cases (with and
without resonances) are very similar, and so are the value of the Poynting
vector and the energy inside the wire. However, the energy outside the wire
presents an additional contribution $W{_o}$ due to the excitons (see inset in
Fig.~\ref{fig:VG}), slowing the propagation of the energy down.

\section{Concluding remarks}

In conclusion, we have shown that that strong coupling between weakly guided
modes of a semiconductor nanowire and a surrounding excitonic medium can be
achieved, exhibiting Rabi splittings of more than $\Omega_R>100\,$meV for an
organic dye. The bare photonic modes are determined through a rigorous classical
analysis, namely, waveguide propagating modes with an evanescent tail outside
the nanowire. The evanescent tail allows for strong coupling of the nanowire
modes with the excitons in external dye molecules. The underlying physical
mechanism is similar to surface plasmon polaritons in metallic nanorods, but
with the advantage of much larger propagating lengths due to the nearly complete
absence of absorption inside the semiconductor nanowire. A quantum model
provides a straightforward analytical expression for the Rabi splitting and
reveals that the relevant quantity is not field concentration, but the fraction
of the field interacting with the emitters. The quantum model also reveals that
coherent energy exchange plays an important role in the coupled system: the
dispersion relations reveal avoided crossings with clear Rabi splittings, as
expected from strong coupling. We furthermore showed that the polariton modes in
these systems can achieve significant propagation lengths up to two orders of
magnitude larger than the bare mode wavelengths, while still maintaining a
significant excitonic character. This happens because strong coupling can shift
exciton modes to frequency regions where material losses are much smaller than
around the exciton resonances. Recalling that lowest-order nanowire guided modes
can be considered nearly 1D lossless propagating
modes~\cite{Paniagua-Dominguez2013}, we thus anticipate that the proposed
configuration might be a suitable candidate for enhanced exciton
conductance~\cite{Feist2015}, which holds promise of applications related to
exciton transport, slow light, and conversion modes.     

\begin{acknowledgments}
The authors acknowledge the Spanish ``Ministerio de Econom\'{\i}a, Industria y
Competitividad'' for financial support through grants MAT2014-53432-C5-5-R and
FIS2015-69295-C3-2-P, the ``Mar\'ia de Maeztu'' programme for Units of
Excellence in R\&D (MDM-2014-0377), and through an FPU Fellowship (D.R.A.) and a
Ramón y Cajal grant (J.F.). We also acknowledge funding from the European
Research Council (ERC-2016-STG-714870).
\end{acknowledgments}

\end{document}